\documentclass[conference]{IEEEtran}
\ifCLASSOPTIONcompsoc
  \usepackage[nocompress]{cite}
\else
\usepackage{cite}
\usepackage{booktabs} 
\ifCLASSINFOpdf
\usepackage[pdftex]{graphicx}
\DeclareGraphicsExtensions{.pdf,.jpeg,.png}
\else
\fi
\usepackage[misc]{ifsym}

\begin{document}
%
\title{BDTF: A Blockchain-Based Data Trading Framework with Trusted Execution Environment}

\author{\IEEEauthorblockN{Guoxiong Su, Wenyuan Yang, Zhengding Luo, Yinghong Zhang, Zhiqiang Bai, Yuesheng Zhu \Letter}\\
\IEEEauthorblockA{School of Electrical and Computer Engineering, Peking University, China\\
Email: suguoxiong@pku.edu.cn, zhuys@pku.edu.cn}
}


%


\maketitle
\begin{abstract}
The need for data trading promotes the emergence of data market. However, in conventional data markets, both data buyers and data sellers have to use a centralized trading platform which might be dishonest. A dishonest centralized trading platform may steal and resell the data seller’s data, or may refuse to send data after receiving payment from the data buyer. It seriously affects the fair data transaction and harm the interests of both parties to the transaction. To address this issue, we propose a novel blockchain-based data trading framework with Trusted Execution Environment (TEE) to provide a trusted decentralized platform for fair data trading. In our design, a blockchain network is proposed to realize the payments from data buyers to data sellers, and a trusted exchange is built by using a TEE for the first time to achieve fair data transmission. With these help, data buyers and data sellers can conduct transactions directly. We implement our proposed framework on Ethereum and Intel SGX, security analysis and experimental results have demonstrated that the framework proposed can effectively guarantee the fair completion of data tradings.

\end{abstract}


%
\IEEEpeerreviewmaketitle

\section{INTRODUCTION}

 
\indent Data has become a key asset in our economy, the utilization of data can bring huge economic benefits or help us make better desicision. For example, by analyzing the data of the user’s product purchase history to generate personalized recommendations, it has brought huge economic benefits to e-commerce companies such as Amazon, Netflix and Alibaba. Using artificial intelligence models trained on large-scale medical data to predict and diagnose diseases has been used to improve the quality of medical services. The performance of these techniques depends heavily on the quality and quantity of data, so the need for expanding the amount of data through data trading arises, which promoted the emergence of data market, such as DataExchange \cite{jung2017accounttrade} and Datacoup \cite{liang2018survey}.\\
\indent A conventional data market includes three main participants, namely: data seller, data buyer and centralized trading platform. In a common trading process, the data seller sends the data to a centralized trading platform and entrusts it to sell. The data buyer browses the information and historical reviews to the data on the centralized trading platform, and select the data of interest to purchase. However, the centralized trading platform may be dishonest and harm the interests of both parties to the transaction. For example, dishonest centralized trading platform may steal and resell the data seller’s data, or may refuse to send data after receiving payment from the data buyer. It poses a huge challenge to conduct a fair data trading.\\
\indent Blockchain is a distributed public ledger based on crypto-graphic technologies first introduced by Nakamoto in 2008 \cite{nakamoto2019bitcoin}. It maintains a continuously growing list of ordered records called blocks. Each block has a hash pointer to the previous block and a Merkle Hash Tree (MHT) is employed to integrate transactions into one block in an efficient way. The advantages of blockchain include: decentralization, immutability and transparency. And these advantages of blockchain paved for fair data trading in data market. \\
\indent In this paper, we design a blockchain-based data trading framework (BDTF), which takes advantages of the blockchain to build a decentralization platform for data trading. In our design, data buyers and data sellers conduct transactions directly on the blockchain, which avoids risks caused by centralized trading platforms. All payment records and reviews generated by data buyers are faithfully recorded in the blockchain through consensus protocol, and can be accessed by every user of BDTF.\\
\indent The decentralized nature of the blockchain has also brought challenge to our work, which makes it difficult to ensure fair payment of data tradings. For example, a malicious data seller may defraud the data buyer by sending fake data, and a malicious data buyer may refuse to pay after receiving the authentic data. To address this important challenge, we first build a trusted exchange by using Trusted Execution Environment (TEE, such as Intel SGX \cite{costan2016intel}, Sanctum \cite{costan2016sanctum} and Keystone \cite{lee2020keystone}) in BDTF to assist in the fair payment of the transactions. During a trading process, the data seller sends data to a trusted exchange first. Then the data buyer can verify whether the data in the trusted exchange is he/she needs. Finally, only when the trusted exchange detected the payment from the data buyer to the data seller will it send the data to the data buyer.\\
\indent The rest of this paper is as follows. Section II briefly introduces the background and related work. The design of the framework is presented in Section III. In Section IV, security and performance analysis are carried out. Finally, this paper is concluded in Section V.

\section{BACKGROUND AND RELATED WORK}
\subsection{Trusted Execution Environment}
\indent A Trusted Execution Environment (TEE) \cite{sabt2015trusted} is a secure area in the processor and it can shield the confidentiality and integrity of the code and data inside from accesses by other software, including higher-privileged software. There are many TEEs have been proposed by both industry and academia, such as Intel SGX \cite{costan2016intel}, Sanctum \cite{costan2016sanctum} and Keystone \cite{lee2020keystone}. In this paper, we use Intel SGX as as the exemplary implementation, but we emphasize that it may use any comparable TEE with attestation capabilities. \\
\indent Intel Software Guard Extensions (SGX) \cite{costan2016intel} provides a widely used TEE implementation for general-purpose computation, which is known as enclaves in SGX. Code running inside a enclaves has a protected address space. When data from a enclave moves off the processor to memory, it is transparently encrypted with keys only available to the processor. Thus the operating system, hypervisor, and other users cannot access the enclave’s memory. In the enclave, the code and data are measured at the startup stage and the measurement is signed into an attestation report based on a hardware-based root of trust. The report can be verified to show the unmodified enclave code logical, by which users can confirm the security of enclave and provision the secret into the enclave at runtime, which is called remote attestation protocol.\\
\indent Due to the powerful functions of SGX, many applications based on SGX have appeared. For example, Schuster et al. \cite{schuster2015vc3} presented VC3, a cloud data analytics framework based on SGX, which allows users to run distributed MapReduce computations in the cloud while keeping their code and data secret. Zhang et al. \cite{zhang2016town} presented Town Crier, which adopts a TEE to enable a secure data-feed for blockchain smart contracts. Cheng et al. \cite{cheng2019ekiden} presented Ekiden, a system uses a TEE to achieve scalable and private smart contracts. Bentov et al. \cite{bentov2019tesseract} proposed Tesseract, a secure muti-blockchain cryptocurrency exchange by using a TEE .

\subsection{Blockchain-based Data Trading}
\indent Recent years have witnessed a surge of studies on blockchain-based data trading. For example, Wang et al. \cite{wang2018novel} proposed a data trading scheme based on the Bitcoin system. In  their scheme, the digital content is encrypted by a symmetric key, and then the symmetric key is encrypted by the RSA scheme. Only the user who paid for the digital content through the Bitcoin system can get the RSA private key for decryption. However, their scheme does not guarantee that the RSA private key provided by the digital content seller is the RSA private key used for encryption.\\
\indent Zhao et al. \cite{zhao2019machine} proposed a blockchain-based fair data trading protocol in big data market, which integrates ring signature, double-authentication-preventing signature and similarity learning to guarantee the availability of trading data, privacy of data providers and fairness of the transaction. However, the market manager in their scheme may collude with data sellers to deceive data consumers.\\
\indent Dib et al. \cite{dib2020novel} proposed a novel framework based on blockchain, which combines a separate secure environment to execute a model on top of data to protect the confidentiality of personal data. However, they assumed that data owners in their system is honest, which is not suitable for most data trading scenarios.\\
\indent Dai et al. \cite{dai2019sdte} proposed SDTE, a blockchain-based data trading ecosystem. In SDTE, buyers of data cannot directly access the original data they purchased, and can only obtains the analysis results of the data, which is generated from Intel SGX. To ensure fair payment, the data from different data sellers will be calculated in different SGX, the result with the most votes will be regarded as the final result, and the data seller who votes the final result is regarded as an honest one and will be rewarded by buyer. However, if there are too many malicious data sellers in a transaction, honest data sellers will not be able to get reasonable compensation.\\
\indent Therefore, it is still a challenge to ensure the fair payment in blockchain-based data trading. This paper is inspired a lot by \cite{dib2020novel} and \cite{dai2019sdte}, the main difference is that they both use TEE to protect the confidentiality of data during data processing. Instead, we use TEE to build a trusted exchange, which is used to achieve fair data transmission.

\section{THE PROPOSED FRAMEWORK}
\indent In this section, we present the proposed data trading framework. We first start by introducing the overview of our framework. We then present notations used in our proposal. Finally, we explain two main workflows of the framework, namely: requirements matching and trading.
\subsection{Overview of BDTF}
\indent As shown in Fig. 1. There are three participants in BDTF, namely: data seller, data buyer and trusted data trading platform (TDTP).

\begin{figure*}[htbp]
\centering
\includegraphics[width=7 in]{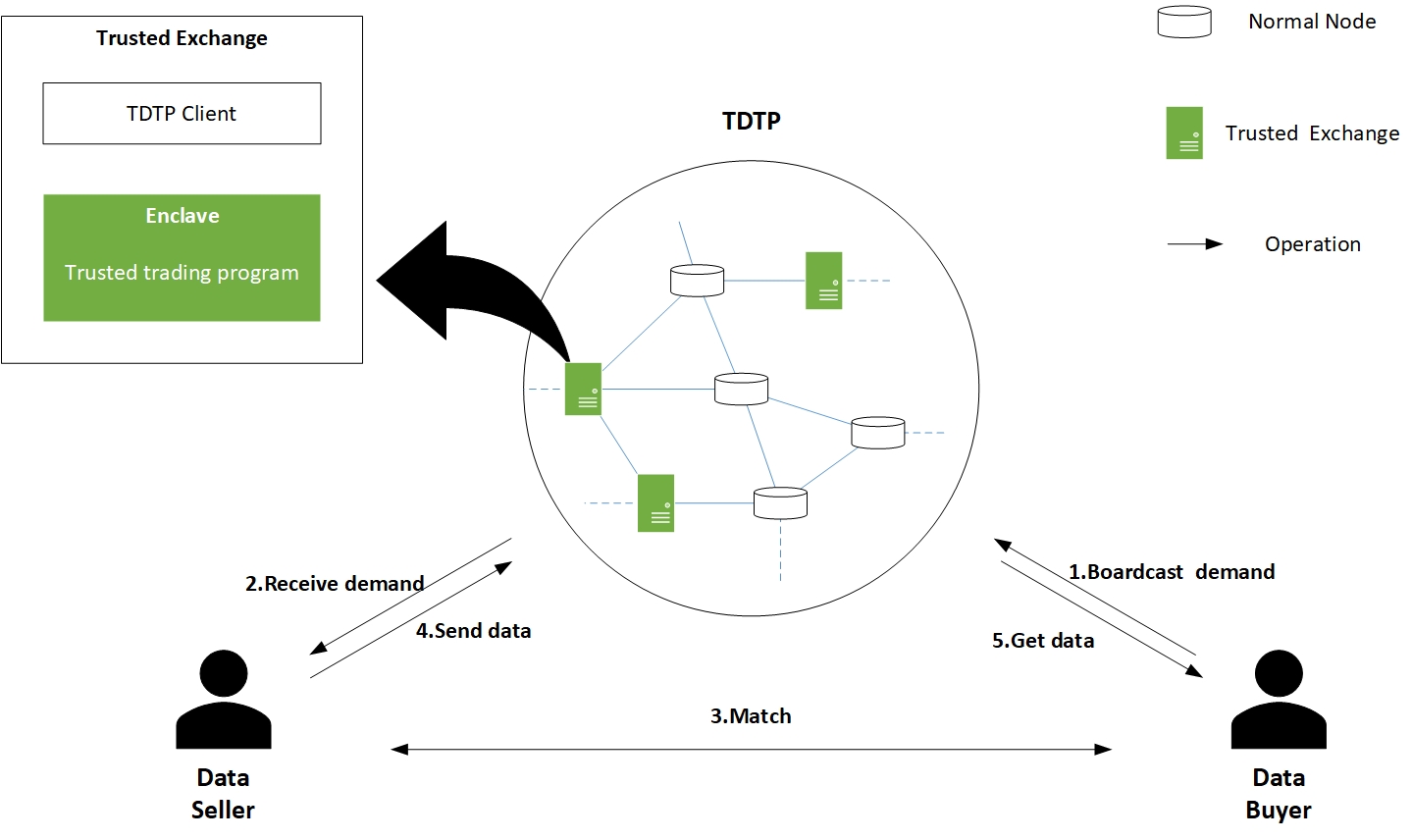}
\centering
\caption{Overview of BDTF}
\label{1}
\end{figure*}
\textbf{Data seller}: A data seller can be any entity that have the right to sell a data item. They can get ether ( For ease of notation, we use ether as the exemplary cryptocurrencies in BDTF) by selling the data they own. 

\textbf{Data buyer}: Data buyers are people or organizations who have a need for data. They broadcast their data demands and complete the purchase of data through a data trading platform.

\textbf{Trusted data trading platform}: Trusted data trading platform (TDTP) is a blockchain network, it consists of two kinds of nodes. The one is normal node, it runs a TDTP client and only participate in the consensus of the blockchain. The other one is trusted exchange, which not only runs a TDTP client for consensus, but also runs a trusted trading program in the enclave, which is used to achieve fair data transmission. In TDTP, each node that supports TEE can become a trusted exchange, and they can make profits by assisting in fair data transmission.\\
\indent In a simplified trading process, the data buyer first boardcasts his/her data demand on TDTP and finds suitable data seller based on the data of interest. After that, the data buyer selects a trusted exchange on TDTP, and informs the data seller the address of the trusted exchange. Then the data seller sends the data to the trusted exchange’s enclave. Finally, only after the trusted trading program running in the trusted exchange’s enclave has detected that the data buyer has paid the data seller can the data buyer receive the data from the enclave. \\
\indent To detect payments from data buyers to data sellers, there is a first-in-first-out (FIFO) queue stored inside the trusted exchange’s enclave, and the size of the queue is set according to a parameter that specifies the maximum time window that the enclave maintains. The enclave receives the latest block headers from the untrusted TDTP client that runs on the same trusted exchange and put these block headers into the FIFO queue at runtime. After that, the trusted trading program will detect payment records 
in the block headers stored in the FIFO queue. \\
\indent To ensure that the detected payment records are authenticity and valid, the enclave verifies the authenticity of each received block header before placing it into the FIFO queue. Specifically, the enclave judges the authenticity of each block header by verifying whether the hash of the previous block stored in the block header is correct and whether the difficulty of the block meets the current difficult level of the consensus protocol rule adopted by TDTP, such as proof-of-work. In addition, the trusted trading program running in the enclave is hardcoded with the hash of the TDTP genesis block, or a more recent “checkpoint” block of the TDTP. These prevent enclave from putting a low-difficulty fake chain to the FIFO queue.

\subsection{Summary of Notations}
A summary of notations used in our proposal is presented in Table I.

\begin{table}[!h]
\caption{Summary of notations}
\label{I}
\centering
\begin{tabular}{l c}
\toprule                             
Notation & Explanation\\
\midrule                          
$A_{role}$ & Blockchain address of role\\
${IP}_{role}$ & The IP address of role\\
$E_{(role1,role2)}$ & Evidence of payment from role1 to role2\\
$K_{role}$ & An AES-256 key created by role\\
ID & Transaction id generated by trusted exchange\\
P & The price of data\\
\bottomrule        
\end{tabular}
\end{table}

\subsection{Workflow}
To describe how the framework works, we introduce two main workflows of the system.
\subsubsection{Requirements Matching}
Before open a trade, the data buyer needs to find appropriate data source through TDTP. We define this process as requirements matching and Fig. 2 shows more detail about it.

\begin{figure}[!h]
\centering
\includegraphics[width=2.5in]{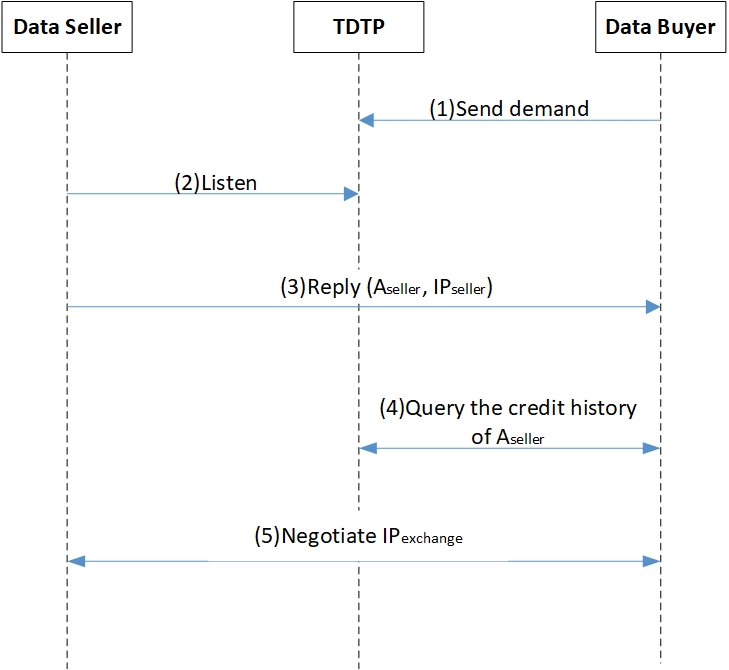}
\caption{The process of requirements matching in TDTP}
\label{2}
\end{figure}

\begin{itemize}
\item Step1: When a data buyer wants to buy data through TDTP, he/she will first broadcast his/her trading demand on TDTP. The broadcast message contains the following: data specifications, P and Contact address of the data buyer (such as IP address ${IP}_{buyer}$).
\item Step 2: Data sellers receive the trading demands broadcasted by data buyers through their own client. They will check that whether the data they own meet the data specifications and whether the bid price is appropriate to decide whether to respond to a trading demand.
\item Step 3: The data seller replys $A_{seller}$ and ${IP}_{seller}$ to the data buyer through the ${IP}_{buyer}$ in the broadcasted message.
\item Step 4: The data buyer selects data seller based on the data of interest and the historical reviews to the data seller, which are recorded in the TDTP.
\item Step 5: The data buyer informs the appropriate data seller to conduct the transaction through ${IP}_{seller}$. Then the data buyer negotiates a trusted exchange for the transaction with the data seller.
\end{itemize}

\subsubsection{Trading}
After requirements matching, the data buyer and data seller have agreed on the trading price and the trusted exchange used for the transaction. The process of trading is shown in Fig. 3.

\begin{figure}[!h]
\centering
\includegraphics[width=3.6 in]{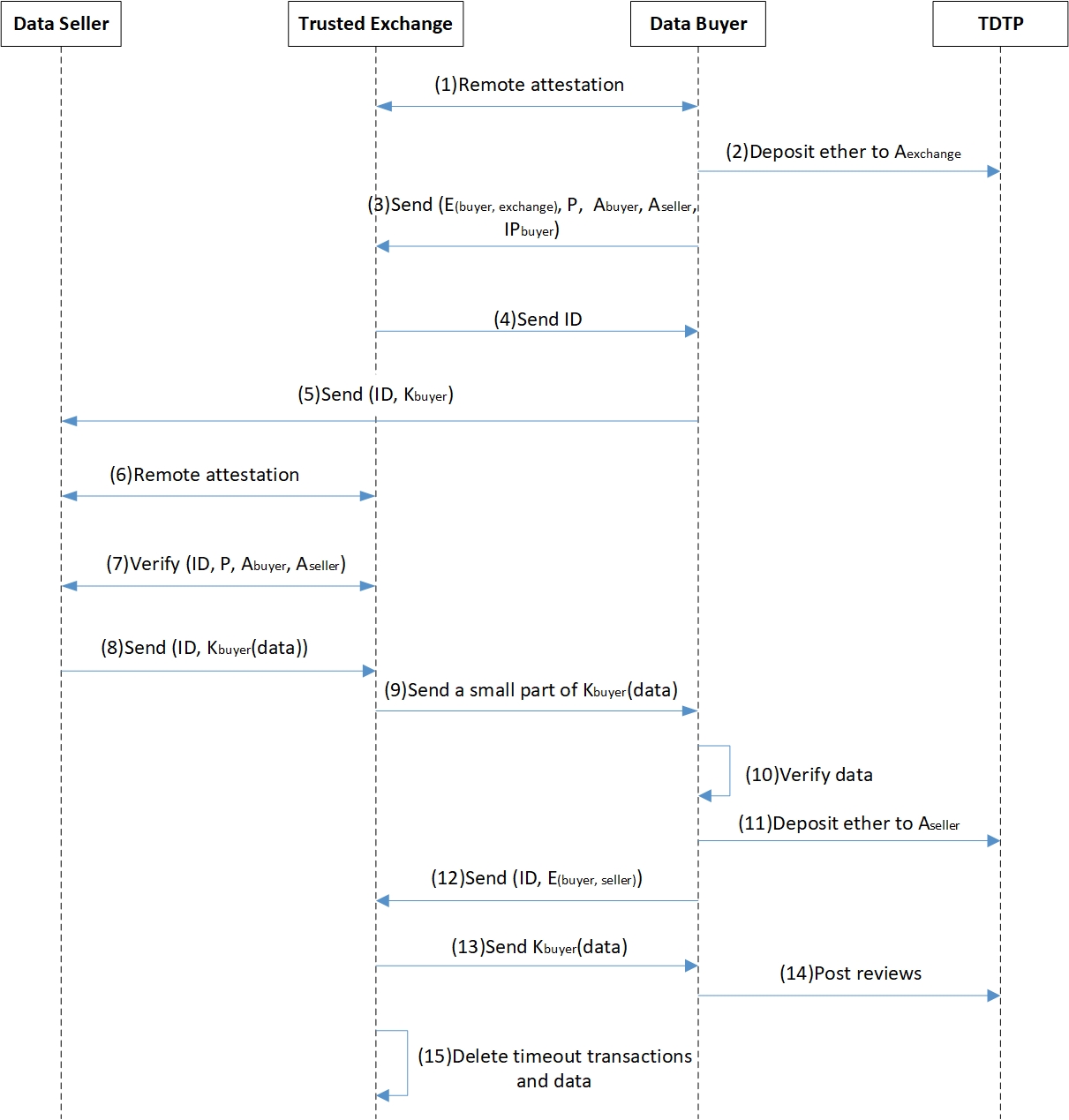}
\caption{The process of trading in TDTP}
\label{3}
\end{figure}

\begin{itemize}
\item Step 1: The data buyer performs remote attestation of the trusted exchange’s enclave to verify that trusted trading program is loaded and run correctly. If the remote attestation result is correct, the data buyer goes to next step.
\item Step 2: The data buyer deposits ether to the blockchain address of the trusted exchange owner $A_{exchange}$ to buy the service provided by the trusted exchange.
\item Step 3: After the deposit transaction is confirmed on the blockchain, the data buyer transforms the confirmed deposit into evidence $E_{(buyer,exchange)}$. $E_{(buyer,exchange)}$ consists of the transaction that spends the ether into the blockchain address of the trusted exchange owner $A_{exchange}$, as well as an authentication path that consists of the sibling nodes in the Merkle tree whose root is stored in a block header, and the index of that block. After that, the data buyer sends $E_{(buyer,exchange)}$, P, $A_{buyer}$ and $A_{seller}$ to the enclave.
\item Step 4: The trusted trading program running in the enclave decides whether to provide service to the data buyer by verifying whether $E_{(buyer,exchange)}$ is valid. If the deposit transaction is valid, which not only means that the enclave detected the deposit transaction in a block, but also that the block containing the deposit transaction is buried under enough additional confirmations, the enclave will generate a transaction id ID, by which the trusted exchange can identify whether a user have permission to use the service, then record ID, P, $A_{buyer}$, $A_{seller}$, ${IP}_{buyer}$ and timestamp of the deposit transaction to a pending transaction table, which be refreshed every once in a while and transactions that have not been completed for a long time will be eliminated. After the above operations are completed, the trusted exchange will send ID to the data buyer through ${IP}_{buyer}$.
\item Step 5: The data buyer sends ID and an AES-256 key $K_{buyer}$, which is used for encrypting data, to the data seller.
\item Step 6: The data seller performs remote attestation of the trusted exchange’s enclave to verify that trusted trading program is loaded and run correctly. If the remote attestation result is correct, the data seller goes to next step.
\item Step 7: The data buyer verifies whether (ID, P, $A_{buyer}$, $A_{seller}$) are set correctly in enclave. If all the parameters are correctly, the data seller will go to step 8.
\item Step 8: The data seller first encrypts the data using $K_{buyer}$, then sends the encrypted data and ID to the enclave.
\item Step 9: After received the data sent by the data seller, the enclave consults the pending transaction table based on the ID provided by the data seller, and sends a small part of the encrypted data to ${IP}_{buyer}$, which is recorded in the pending transaction table. 
\item Step 10: The data buyer decrypts the data received and verifies whether the data is what he/she wants. If the data buyer finds that the data does not match, he/she can terminate the transaction at this step.
\item Step 11: The data buyer deposits ether to $A_{seller}$.
\item Step 12: After the deposit transaction is confirmed on the blockchain, the data buyer sends ID and $E_{(buyer,seller)}$ to the enclave.
\item Step 13: The enclave verifies that the deposit transaction is valid, and checks whether P, $A_{buyer}$ and $A_{seller}$ in the deposit transaction are consistent with the pending transaction table. If everything is correct, the enclave will send data to the data buyer through the ${IP}_{buyer}$ .
\item Step 14: After the data purchase is completed, the data buyer can post review to the data seller according to the quality of the data, and can also post review to the trusted exchange owner according to the services provided by the trusted exchange. Due to the immutability of the blockchain, all reviews will be faithfully recorded on TDTP, later users will be able to refer to these information to select data sellers and trusted exchanges.
\item Step 15: The enclave deletes completed or timeout transactions from the pending transaction table according to the timestamp recorded in the table, and the data related to these transactions will also be deleted.
\end{itemize}

\section{Security and Performance Analysis}
In this section, we conduct a security and performance analysis of the entire framework. We first analyze the possible attack of every malicious participant. Then a performance evaluation is performed.
\subsection{Security Analysis}
\subsubsection{Malicious data seller} A malicious data seller may sell fake or mismatched data to the data buyer. To solve this problem, the data buyer can view the historical trading reviews to the data seller on TDTP to decide whether to trade with the data seller. In addition, after the data buyer sends the data to the enclave, the data buyer can view a small part of the data to verify whether the data provided by the data seller is what he/she wants. If the data seller provides fake or mismatched data, the data buyer can refuse payment.\\ 
\indent Finally, after the transaction is completed, the data buyer can give review to the data seller on the TDTP. The malicious data seller will receive a bad review, which will affect his/her future earnings.

\subsubsection{Malicious data buyer} A malicious data buyer may refuse to pay after receiving the authentic data. However, in our solution, the trusted trading program running in the trusted exchange’s enclave will only send data to the data buyer when it detects that the data buyer's payment to the data seller has been confirmed.\\
\indent A malicious data buyer may also launch a double-spend attack. In order to deal with this possible attack, the enclave will only consider a payment record valid after it is buried under enough additional confirmations.

\subsubsection{Malicious trusted exchange} Our scheme requires that the source code of the trusted trading program running in the enclave is open source, so we assume that the trusted trading program is trusted. However, after the data buyer paid to the data seller, a malicious trusted exchange may shut down and stop working, which will prevent the data buyer from getting the data. To avoid this risk, the data buyer and data seller can choose multiple trusted exchanges to conduct a trading. In this case, when the data seller sends data, he/she needs to send the same data to multiple enclaves. The data buyer only need to pay once can he download data from these enclaves, which greatly reduces the risk of single point of failure/attack. \\
\indent Besides, a malicious trusted exchange may feed a low-difficulty fake chain to the enclave, the solution to this attack has been introduced in Section III. \\
\indent After the transaction is completed, the data buyer can also give review to the service provided by the trusted exchange, and the malicious trusted exchange will receive a bad review, which will affect it’s future earnings.

\subsection{Performance Analysis}
As described in the previous section, the TDTP composed of two kinds of node, namely normal node and trusted exchange. The TDTP client running on both nodes constitutes a blockchain. Although in addition to the TDTP client, an enclave is running on the trusted exchange, it does not affect the performance of the blockchain. Therefore, to analyze the performance of the whole solution, it suffices to separately analyze the performance of the blockchain and the trusted exchange .\\
\indent To analyze the performance of the blockchain, we took Ganache $\footnote{https://github.com/trufflesuite/ganache}$, a local and virtual Ethereum blockchain for testing, as the blockchain base. And we used web3.js, a popular library for JavaScript that allows programmers to interact with the Ethereum blockchain, to randomly send ether transfer transactions and reviews from one account to another. We tested the transactions throughput and the validation latency of the blockchain. The former means the number of transactions that the blockchain can validate per second, and the latter means the time required to validate one single transaction. These two indicators are analyzed versus the frequency of transactions that are sent to the blockchain. To ensure the accuracy of the results, our experiment was repeated 100 times.\\
\indent The results in Fig. 4 indicate that number of transactions validated per second is impacted by the transactions frequency. Up to 300 transactions per second, the throughput is optimal. When the throughput is greater than 300, the average throughput begins to lag behind the frequency of transactions.\\
\indent The results in Fig. 5 show that the time required to validate a transaction increases as the frequency of transactions increases. Due to the throughput lags behind the frequency of transactions, the maximum latency increases a lot, after the frequency of transactions is greater than 300.
 \begin{figure}[h]
\centering
\includegraphics[width=3 in]{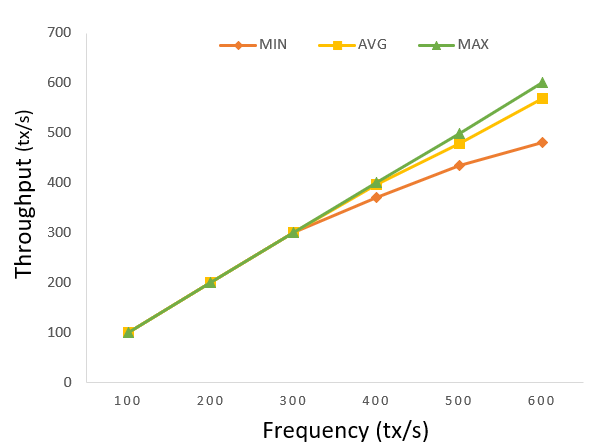}
\caption{The throughput of blockchain}
\label{4}
\end{figure}

\begin{figure}[h]
\centering
\includegraphics[width=3 in]{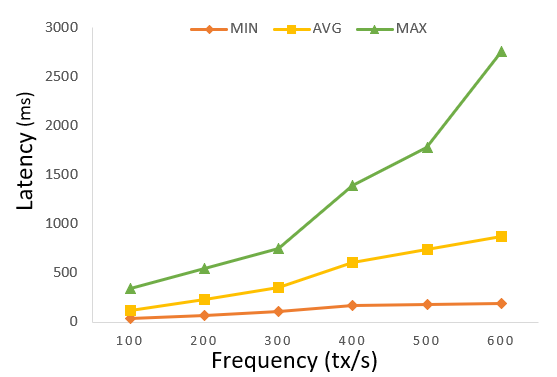}
\caption{The latency of blockchain}
\label{5}
\end{figure}
\indent We also implemented a prototype of trusted exchange on a desktop with an Intel i7-7700 CPU and 16GB memory. We measured the enclave response time for handing service requests from data buyers. The experiment was repeated 1000 times, and we found that the trusted exchange can respond to requests quickly, the average time we got for the process is 42ms, 38ms in the best case and 47 ms in the worst case.

\section{Conclusion}
In this paper, we presented the design of BDTF, a blockchain-based decentralized framework for data trading. We analyzed that the conventional centralized data trading platform may be dishonest and harm the interests of both parties to the transaction. We formalized BDTF to address this issue. In our solution, we used blockchain to allow data buyers and data sellers conduct transactions directly. And we built a trusted exchange by using Intel SGX to achieve fair data transmission for data trading. Besides, we analyzed the security and performance of the proposed framework, the results demonstrated that our framework can effectively guarantee the fair completion of data tradings.


\section*{Acknowledgment}
This work was supported in part by NSFC-Shenzhen Robot Jointed Founding under Grant U1613215, in part by the Shenzhen Municipal Development and Reform Commission (Disciplinary Development Program for Data Science and Intelligent Computing), and in part by the Key-Area Research and Development Program of Guangdong Province under Grant 2019B010137001.



%

\end{document}